\documentclass[10pt,onecolumn]{article}

\usepackage{amsmath}
\usepackage{amsfonts}
\usepackage{amssymb}
\usepackage{graphicx}
\usepackage{grffile}
\usepackage{braket}
\usepackage{subcaption}
\usepackage{stfloats}
\usepackage{listings}
\usepackage{bm}
\usepackage{qcircuit}
\usepackage{xcolor}
\usepackage{authblk}
\usepackage[hidelinks]{hyperref}

\title{Optimization strategies in WAHTOR algorithm for quantum computing empirical ansatz: a comparative study}
 
\author[1,2]{Leonardo Ratini}
\author[2]{Chiara Capecci}
\author[1]{Leonardo Guidoni
\thanks{leonardo.guidoni@univaq.it}}
\date{}
\affil[1]{Dipartimento di Scienze Fisiche e Chimiche, Università degli Studi dell'Aquila, Coppito, L'Aquila, Italy}
\affil[2]{Dipartimento di Ingegneria e Scienze dell'Informazione e Matematica, Università degli Studi dell'Aquila, Coppito, L'Aquila, Italy}

\begin{document}

\maketitle
\begin{abstract} 
By exploiting the invariance of the molecular Hamiltonian by a unitary transformation of the orbitals it is possible to significantly shorter the depth of the variational circuit in the Variational Quantum Eigensolver (VQE) algorithm by using the Wavefunction Adapted Hamiltonian Through Orbital Rotation (WAHTOR) algorithm. This work introduces a non-adiabatic version of the WAHTOR algorithm and compares its efficiency with three implementations by estimating Quantum Processing Unit (QPU) resources in prototypical benchmarking systems. 
Calculating first and second-order derivatives of the Hamiltonian at fixed VQE parameters does not introduce a significant QPU overload, leading to results on small molecules that indicate the non-adiabatic Newton-Raphson method as the more convenient choice. On the contrary, we find out that in the case of Hubbard model systems the trust region non-adiabatic optimization is more efficient. The preset work therefore clearly indicates the best optimization strategies for empirical variational ansatzes, facilitating the optimization of larger variational wavefunctions for quantum computing.
\end{abstract}

\section{INTRODUCTION}
Quantum computers and algorithms have brought about significant advancements in computational capabilities, potentially revolutionising numerous scientific fields. \cite{Shor1997,Bernien2017,Banuls2020,Hussain2020,Egger2021} Leveraging the principles of quantum mechanics, these technologies enable the manipulation and processing of information in ways that are fundamentally different from classical computers. \cite{Preskill2018, Arute2019} Among the many applications of quantum algorithms, quantum simulations \cite{Stair2021,Barison2021} stand out as a field that stands to benefit significantly from these advancements, particularly in quantum chemistry. \cite{Cao2019,Bauer2020,McArdle2020} 
One of the leading quantum algorithms for addressing quantum chemistry problems is the Variational Quantum Eigensolver (VQE). \cite{Peruzzo2014, McClean2016, Kandala2017,Fedorov2022,Tilly2022} VQE combines classical and quantum resources to find the ground state energy of a molecular system. By employing a parameterized quantum ansatz representing a trial wave function, VQE iteratively adjusts the parameters to minimize the energy through measurements performed on a quantum computer. This hybrid approach allows for exploring the complex quantum landscape associated with molecular systems.
Despite its promise, VQE faces challenges in scaling to larger systems due to the exponential growth of the parameter space and, more generally, due to the implementation of the longer circuits describing wavefunctions on noisy devices.
To overcome these obstacles, many algorithms have been developed to improve the performance of the VQE. \cite{Benfenati2021,Barkoutsos2018,Grimsley2019,Yordanov2021,Castaldo2022} Moreover using heuristic ans\"atze \cite{Ganzhorn2019,Tkachenko2020}, rather than the chemically inspired ones \cite{Kutzelnigg1991, Romero2019, Xia2020}, allows us to consider more shallow circuits. 
Another solution being explored is the development of pulse-based ans\"atze, which aim to optimize and shorten circuit lengths.  \cite{Egger2023,Meitei2021,Liang2022}
One of the methods that have emerged as a potential solution is the Wave Function Adapted Hamiltonian Through Orbital Rotations (WAHTOR). \cite{Ratini2022}
This algorithm optimizes molecular orbitals by adapting the Hamiltonian to the chosen topology of the wave function ansatz, utilizing orbital rotations to facilitate more efficient and accurate simulations within the VQE framework.  
Essentially, the algorithm works as follows: once defined the initial Hamiltonian $H$ in the Hartree-Fock basis, we perform a first VQE algorithm. Then we apply to the initial Hamiltonian a parameterized unitary operator $\hat{U_1}=\hat{U_1}(\bm{R_1})$, corresponding to a change of the single-particle basis set. The vector of the rotation parameters $\bm{R_1}$ is chosen in order to find the operator $\hat{U_1}$ that reduces the expectation value of the energy. 
The transformed Hamiltonian $\hat{U_1}^{\dag}(\bm{R_1})H\hat{U_1}(\bm{R_1})$ is used to perform a VQE optimization for the re-optimisation of the wave function. Once the ansatz optimisation is converged
we determine, keeping the ansatz parameters fixed, a new operator $\hat{U_2}=\hat{U_2}(\bm{R_2})$, corresponding to an additional rotation that is applied to the Hamiltonian used in the previous VQE, so the resulting unitary transformation is obtained by applying the operator $\hat{U_1}(\bm{R_1})\hat{U_2}(\bm{R_2})$ to the initial Hamiltonian to perform another VQE.
We alternate a VQE run and the definition of a new Hamiltonian until energy convergence is reached.
Thus the final Hamiltonian can be calculated by applying the operator given by the product of the single unitary operators $\hat{U}=\hat{U_1}(\bm{R_1})\hat{U_2}(\bm{R_2})\hat{U_3}(\bm{R_3})\dots$ to the initial Hamiltonian, i.e. the one expressed in the Hartree-Fock basis. 
Orbital optimization in the context of the VQE quantum algorithm is a method also adopted in the ref. \cite{Mizukami2020}: in this work, the optimization exploits derivatives up to the second order and the optimizer moves along the direction of the Newton method. 
On the other end, the WAHTOR method is a useful approach that offers flexibility in choosing optimizers capable of leveraging the analytic derivatives of any order, as shown in the present work. 
Optimizers play a crucial role in the process, aiming to identify the optimal set of parameters that minimize the energy of a given system. \cite{Nocedal1999,Fletcher1987} Different optimizers employ a variety of strategies and algorithms that efficiently explore the energy landscape, resulting in faster convergence and improved accuracy.
In this study, we compare the original algorithm, called adiabatic, with a new procedure, called non-adiabatic that we will illustrate in section \ref{results}.
The main focus is to explore the potential of the non-adiabatic versions of the WAHTOR algorithm.
Through the analysis of these optimizers, our objective is to examine their convergence behaviour, computational efficiency and accuracy. This evaluation aims to enhance our understanding of their functioning and contributions to the optimization of molecular orbitals in quantum chemistry simulations.
It is important to note that the WAHTOR method extends beyond molecular systems, finding relevance in other quantum systems such as the Hubbard model. \cite{Sokolov2020a,Cade2020,Suchsland2022,Consiglio2022}
By exploring the performance of these optimizers on different systems, we can gain insight into their versatility and potential applications of the algorithm to different quantum systems.
In section \ref{computational_details}, we provide the computational specifics of the algorithm, including the systems studied, the ansatz used, and a description of the optimization strategies employed. In section \ref{algorithm_results}, we present the mathematical treatment of the non-adiabatic algorithm and we analyze the performance of the optimization methods for each system considered. Specifically, we examine the optimization steps in relation to the number of measured strings, which reflects the utilization of QPU resources during the optimization. Lastly, in section \ref{section:conclusions}, we discuss the results obtained.
\section{COMPUTATIONAL DETAILS}
\label{computational_details}

\begin{figure}[h]
\centering
\begin{equation}
\Qcircuit @C=0.7em @R=0.3em {
& \lstick{\ket{0}} & \gate{R_y(\theta_0)} & \qw & \ctrl{1} & \qw & \qw & \gate{R_y(\theta_8)} & \qw & \ctrl{4} & \qw  & \qw & \qw &  \gate{R_y(\theta_{16})} & \qw &\\
& \lstick{\ket{1}} & \gate{R_y(\theta_1)} & \qw & \targ & \ctrl{1} & \qw & \gate{R_y(\theta_{9})} & \qw & \qw & \ctrl{4} & \qw &  \qw &\gate{R_y(\theta_{17})} & \qw &\\
& \lstick{\ket{2}} & \gate{R_y(\theta_2)} & \qw & \qw & \targ &  \ctrl{1} & \gate{R_y(\theta_{10})} & \qw & \qw & \qw & \ctrl{4} & \qw  &\gate{R_y(\theta_{18})} & \qw &\\
& \lstick{\ket{3}} & \gate{R_y(\theta_3)}& \qw  & \qw & \qw & \targ & \gate{R_y(\theta_{11})} & \qw & \qw & \qw & \qw  & \ctrl{4} &  \gate{R_y(\theta_{19})}  & \qw &\\
& \lstick{\ket{4}} & \gate{R_y(\theta_4)} & \qw  & \ctrl{1} & \qw & \qw & \gate{R_y(\theta_{12})} & \qw & \targ & \qw & \qw & \qw & \gate{R_y(\theta_{20})}  & \qw &\\
& \lstick{\ket{5}} & \gate{R_y(\theta_5)} & \qw & \targ & \ctrl{1} & \qw & \gate{R_y(\theta_{13})} & \qw & \qw & \targ & \qw & \qw  & \gate{R_y(\theta_{21})}  & \qw &\\
& \lstick{\ket{6}} & \gate{R_y(\theta_6)}& \qw  & \qw & \targ &  \ctrl{1} & \gate{R_y(\theta_{14})} & \qw & \qw & \qw & \targ & \qw & \gate{R_y(\theta_{22})} & \qw &\\
& \lstick{\ket{7}} & \gate{R_y(\theta_7)} & \qw & \qw & \qw & \targ & \gate{R_y(\theta_{15})} & \qw & \qw & \qw & \qw & \targ  & \gate{R_y(\theta_{23})} & \qw & \\
\ar@{-}[]+<5.3em,15.8em>;[]+<5.3em,-0.0em>
\ar@{-}[]+<5.4em,15.8em>;[]+<5.4em,-0.0em>
\ar@{-}[]+<15.5em,15.8em>;[]+<15.5em,-0.0em>
\ar@{-}[]+<15.6em,15.8em>;[]+<15.6em,-0.0em>
\ar@{-}[]+<26.7em,15.8em>;[]+<26.7em,-0.0em>
\ar@{-}[]+<26.8em,15.8em>;[]+<26.8em,-0.0em>
}
\notag
\end{equation}
\caption{Entangler map for the Hubbard system.}
\label{fig:hubburd_ansatz}
\end{figure}
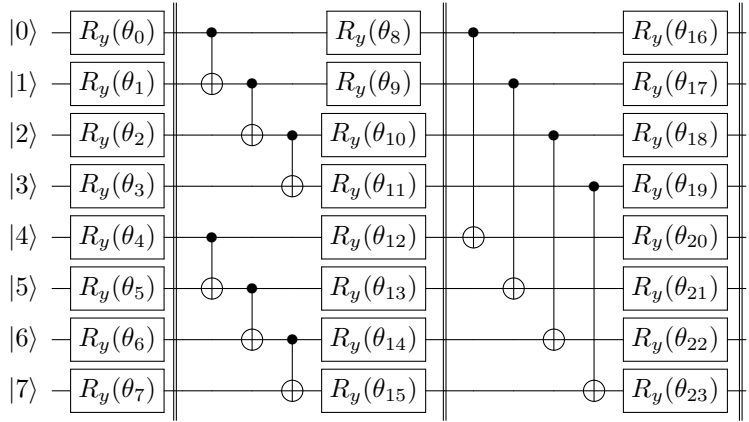

{\it \textbf{Implementation of derivatives}}\\
For the implementation of the WAHTOR algorithm, the matrices responsible for the basis change consider spin symmetry. This means that the spin-up and the spin-down single-particle states are not linearly combined and they are transformed using the same generic unitary transformation, represented by the matrix $U$. This results in a set of $\frac{n^2+n}{2}$ symmetric and $\frac{n^2-n}{2}$ antisymmetric matrices as generators for a system with $n$ single-particle states for each spin projection. In the case of molecules, by linearly combining only spin-orbitals with the same spatial symmetry, we can reduce the number of variables (or parameters) in the Hamiltonian lowering the computational cost.\\
{\it \textbf{Simulated systems}}\\
We tested the optimization methods of the WAHTOR algorithm on two molecular systems and on the Hubbard model, which represent an example of a strongly correlated system. For the molecule, the Hamiltonian has been calculated using the PySCF Python package.\cite{pyscf} The systems studied, along with their descriptions, are as follows:
\begin{itemize}
  \item Hydrogen fluoride ($HF$): atoms were set at the bond distance of $0.917$\AA{} using the 'sto-3g' basis set. The system, obtained through the Jordan-Wigner encoding method and the frozen core approximation, comprises 10 qubits. For simulations, a heuristic ansatz with 2 blocks was employed. Each block includes a layer of rotations around the y-axis on each qubit, as well as a simple ladder map of entanglers (CNOT gates) in which each qubit is the target of the previous qubit and the control of the following qubit. 
  \item Water molecule ($H_2O$): the atoms are arranged to form an isosceles triangle with oxygen positioned at the vertex, the distance between the oxygen and the hydrogen is equal to $0.957$\AA{} while the angle at the vertex, between the two equal sides of the triangle, is equal to $104,5^{\circ}$. The 'sto-3g' basis set was utilized. The resulting system, obtained using the Jordan-Wigner mapping \cite{Jordan1993} and the frozen core approximation, consists of 12 qubits. The blocks of the ansatz are the same used for the $HF$ molecule but with a depth of 4 instead of 2.
  \item 4-site ring Hubbard model in the half-filling regime: the Hamiltonian of the system is:
  \begin{equation}
    H=\sum_{<i,j>}\sum_{\sigma=\uparrow,\downarrow}-\big(a_{i\sigma}^{\dag}a_{j\sigma}+a_{j\sigma}^{\dag}a_{i\sigma}\big)+V\sum_{i=1}^4 n_{i\uparrow}n_{i\downarrow}+\mu \sum_i^4(n_i-2)^2
\label{eq:hubbard}
\end{equation}
    where $n_{i\sigma}=a_{i\sigma}^{\dag}a_{i\sigma}$ with $\sigma=\{\uparrow,\downarrow\}$, $V$ is the on-site potential, $n_i=n_{i\uparrow}+n_{i\downarrow}$, $\mu$ is the chemical potential preserving the requested number of particle for each spin and $<i,j>$ denote the nearest-neighbour. We use the value $V=8$ and $\mu=8$, which is a strongly correlated regime.
    We applied the Jordan-Wigner mapping obtaining an 8-qubit system and the heuristic ansatz considered is composed of 7 blocks: each odd block consists of a pattern of entanglers given by the map $[0,1],[1,2],[2,3],[4,5],[5,6],[6,7]$ and each even block by the map $[0,4],[1,5],[2,6],[3,7]$. The first two blocks of the ansatz are shown in figure \ref{fig:hubburd_ansatz}.
\end{itemize}

{\it \textbf{Ansatz optimization}}\\
The ansatz parameters are represented with the vector $\bm{\theta}$, initialized randomly in the range $[0,2\pi)$. For all the simulated systems, the VQE algorithm optimizes these parameters using the BFGS optimizer in the state vector mode and the convergence threshold is set to $10^{-6}Eh$. The Qiskit library \cite{qiskit} has been used to prepare the quantum circuit, map the fermionic system into qubits and evaluate the cost function. 

{\it \textbf{Hamiltonian optimization strategies}}\\
Two optimization strategies have been considered for the WAHTOR algorithm: the first is the adiabatic optimization, as described in ref. \cite{Ratini2022} and the second is the non-adiabatic optimization, illustrated in the next section. 
The distinction between adiabatic and non-adiabatic methods lies in the variational quantum state that is taken into account during any calculation of the energy derivatives with respect to the Hamiltonian parameters. In the first case, according to the Hellmann-Feynman theorem, the parametrized quantum state is the one that minimizes the energy functional with respect to that value of the Hamiltonian parameters. \cite{Ratini2022} For the non-adiabatic method, the energy is simply a function of the Hamiltonian parameters so that the resulting quantum state, in general, is not at its variational minimum or, if it is, 
the energy derivatives of order greater than one can be calculated as the expectation value of the Hamiltonian derivatives operator. 
Derivatives calculation requires classical and quantum resources, for this reason we implemented three different optimization strategies of the Hamiltonian for the non-adiabatic algorithm:
\begin{itemize}
    \item non-adiabatic trust region: the cost function is approximated with a quadratic model into a region around the current point. The radius of this trust region is determined at each step and the optimum value of the cost function is determined inside this region. \cite{Nocedal1999} In each step of the Hamiltonian optimization routine, the first and second derivatives of the cost function are determined using the equations described in the next section. The new Hamiltonian is found and another step of this routine start. The cycle continues until convergence is reached.
    \item non-adiabatic Newton-Raphson: in the Hamiltonian optimization routine, the $\hat{U}(\bm{R})$ operator is determined by taking
    \begin{equation}
       \bm{R}=-(\nabla^2E(\bm{R},\bm{\theta})|_{\bm{R}=\bm{0}})^{-1}\nabla E(\bm{R},\bm{\theta})|_{\bm{R}=\bm{0}} 
    \end{equation}
    following the idea of the Newton-Raphson optimizer \cite{Ypma1995}. The operator $\hat{U}(\bm{R})$ give us the new Hamiltonian.
    \item non-adiabatic BFGS: this method exploits the BFGS (Broyden-Fletcher-Goldfarb-Shanno) optimization algorithm to update the cost function. \cite{Broyden1970, Fletcher1970, Goldfarb1970, Shanno1970} 
    A quadratic model of the objective function is built but, in this case, the Hessian matrix is approximated using the incremental ratio of the gradient. In our implementation, this method does not use derivatives of the cost function so even the gradient is approximated. In this case, the derivatives are not calculated but their estimation is repeated until convergence is reached. 
\end{itemize}
The three strategies illustrated above and the adiabatic strategy have been compared to evaluate the convergence behaviour and assess their advantages and limitations in terms of Quantum Processing Unit (QPU) resources. We remark that the QPU resources are not directly linked to the number of optimization steps for the Hamiltonian since also the number of optimization steps for the $\bm{\theta}$ parameters of the wavefunction depends on whether the optimization has been carried out adiabatically or non-adiabatically.
We, therefore, assume as a good estimation of the QPU resource directly the number of Pauli strings evaluations performed on quantum computing units.\\

\section{RESULTS}
\label{algorithm_results}
\subsection{Non-adiabatic WAHTOR algorithm}
\label{algorithm}
In this subsection, we illustrate in detail the non-adiabatic WAHTOR algorithm. Note that the notation is different from the one used in \cite{Ratini2022}.
The idea behind this method is to transform the Hamiltonian with the aim to adapt it to a given empirical ansatz. To this aim, we can exploit the property that the eigenvalues of the Hamiltonian $H$ 
\begin{equation}
H=\sum_{i,j}h_{ij}a^\dag_i a_j+\frac{1}{2}\sum_{c,d,e,f}g_{cdef}a^\dag_c a^\dag_d a_e a_f 
\label{eq:H}
\end{equation}
and the eigenvalues of $\hat{U}^{\dag}H\hat{U}$ are the same if $\hat{U}$ is a unitary operator. 
Here we introduced a set of $N$ single-particle states with the corresponding creation and the annihilation operators, respectively $a_i^\dag$ and $a_i$ for $i=1,\dots,N$. 
For the unitary operator, we choose the following parametrized form:
\begin{equation}
   \hat{U}=\hat{U}(\bm{R})=e^{\sum_{jk}(i\bm{T}\cdot\bm{R})_{jk}a_k^{\dag}a_k}
\end{equation}
where $\bm{T}$ and $\bm{R}$ are two vectors, the components of the former are the generators of the unitary matrix's Lie group with dimension $NxN$, whereas the latter is a vector of real components that are the variational parameters corresponding to the specific transformation applied to the Hamiltonian. 
The new operators $b_i^{\dag}=\hat{U}a_i^{\dag}\hat{U}^{\dag}$ and $b_i=(b_i^{\dag})^{\dag}$ satisfy the fermionic anticommutation relations
\begin{equation}
    \{b_i^{\dag},b_j^{\dag}\}=0 \qquad \{b_i,b_j^{\dag}\}=\delta_{ij} \qquad i,j=1,\dots,N
\end{equation}
It follows that we can interpret these operators as the ones associated with a different set of $N$ single-particle states that can be used to describe our quantum system.
So, by exploiting the relation 
\begin{equation}
   b_k^{\dag}=\sum_{j=1}^n U(\bm{R})_{jk}a_j^{\dag} \qquad U(\bm{R})=e^{i\bm{T}\cdot\bm{R}}
\end{equation}
the transformed Hamiltonian $\hat{U}^{\dag}H\hat{U}$ can be expressed in the initial single-particle basis set as
\begin{equation}
H(\bm{R})\equiv\hat{U}^{\dag}(\bm{R})H\hat{U}(\bm{R})=\sum_{i,j}h(\bm{R})_{ij}a^\dag_i a_j+\frac{1}{2}\sum_{c,d,e,f}g(\bm{R})_{cdef}a^\dag_c a^\dag_d a_e a_f 
\label{eq:transformedH}
\end{equation}
where $h(\bm{R})$ and $g(\bm{R})$ are the transformed tensors of respectively $h$ and $g$
\begin{align}
   & h(\bm{R})=e^{-i\bm{T}\cdot\bm{R}}h e^{i\bm{T}\cdot\bm{R}} \\
   & g(\bm{R})=e^{-i\bm{T}\cdot\bm{R}}\otimes e^{-i\bm{T}\cdot\bm{R}}g e^{i\bm{T}\cdot\bm{R}}\otimes e^{i\bm{T}\cdot\bm{R}} 
\end{align}
Now, we represent with $\ket{A_i}$ and $\ket{B_i}$, for $i=1,\dots,2^N$, the Slater determinants obtained by applying on the vacuum state $\ket{\varnothing}$ the operators belonging respectively to the sets $\{a^{\dag}\}$ and $\{b^{\dag}\}$. It follows that for every parametrized state $\ket{\Psi(\bm{\theta})}=\sum_{i=1}^{2^N}c_i(\bm{\theta})\ket{A_i}$, the application of the operator $\hat{U}$ give the new ket
\begin{equation}
    \ket{\Phi(\bm{\theta})}=\hat{U}\ket{\Psi(\bm{\theta})}=\sum_{i=1}^{2^N}c_i(\bm{\theta})\ket{B_i} \qquad c_i(\bm{\theta}) \in \mathbb{C} \qquad i=1,\dots,2^N
\end{equation}
Since the coefficients of the linear combination do not change, $\ket{\Psi(\bm{\theta})}$ or $\ket{\Phi(\bm{\theta})}$ can be represented by a quantum circuit with the same value of the $\bm{\theta}$ parameters by mapping on the qubits the single-particle states corresponding respectively to the set $\{a^{\dag}\}$ or $\{b^{\dag}\}$.
The energy cost function
\begin{equation}
E(\bm{R},\bm{\theta})=\bra{\Psi(\bm{\theta})}H(\bm{R})\ket{\Psi(\bm{\theta})}
\label{eq:energy}
\end{equation}
depends on both set of parameters $\bm{\theta}$ and $\bm{R}$.
\begin{figure}[h!]
\includegraphics[width=1.0\linewidth]
{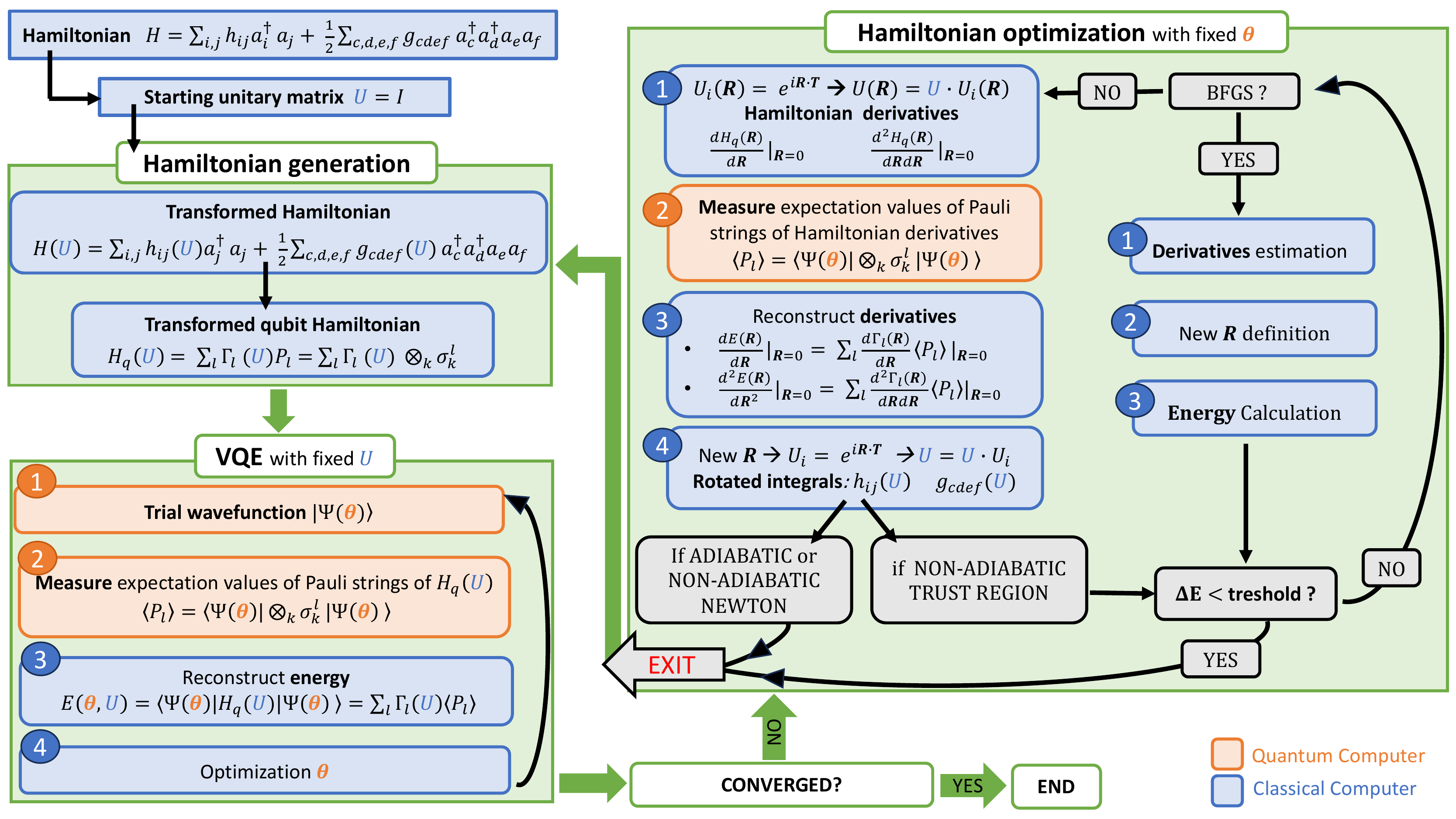}
\caption{The scheme of the WAHTOR algorithm for both the optimization strategies, adiabatic and non-adiabatic defined in this article.}
\label{fig:scheme}
\end{figure}
Energy can be optimized with respect to $\bm{R}$ parameters by calculating its n-order analytical derivatives:
\begin{equation}
\begin{split}
\frac{d^n E(\bm{R},\bm{\theta})}{d \bm{R}^n}=\sum_{i,j}\frac{d^n h(\bm{R})_{ij}}{d \bm{R}^n}\bra{\Psi(\bm{\theta})}a^\dag_i a_j\ket{\Psi(\bm{\theta})} \\
+\frac{1}{2}\sum_{c,d,e,f}\frac{d^n g(\bm{R})_{cdef}}{d \bm{R}^n}\bra{\Psi(\bm{\theta})}a^\dag_c a^\dag_d a_e a_f\ket{\Psi(\bm{\theta})}
\end{split}
\label{eq:dH}
\end{equation}
from which one can see the dependence on the n-order derivatives of $h(\bm{R})$ and $g(\bm{R})$. In particular, we can write down the first and second-order derivatives in $\bm{R}=\bm{0}$, that will be used along the optimization:
\begin{equation}
\begin{split}
& \frac{dh(\bm{R})}{dR_{l_1}}|_{\bm{R}=\bm{0}}=i[T_{l_1},h] \\
& \frac{dg(\bm{R})}{dR_{l_1}}|_{\bm{R}=\bm{0}}=i[T_{l_1}\otimes I+I\otimes T_{l_1},g]
\end{split}
\label{eq:first_H}
\end{equation}
and 
\begin{equation}
\begin{split}
&\frac{d^2 h(\bm{R})}{dR_{l_1}dR_{l_2}}|_{\bm{R}=\bm{0}}=-\frac{1}{2}([T_{l_1},[T_{l_2},h]]+[T_{l_2},[T_{l_1},h]]) \\
&\frac{d^2 g(\bm{R})}{dR_{l_1}dR_{l_2}}|_{\bm{R}=\bm{0}}=-\frac{1}{2}([T_{l_1}\otimes I+I\otimes T_{l_1},[T_{l_2}\otimes I+I\otimes T_{l_2},g]]+ \\
& +[T_{l_2}\otimes I+I\otimes T_{l_2},[T_{l_1}\otimes I+I\otimes T_{l_1},g]])
\end{split}
\label{eq:second_H}
\end{equation}
where $I$ is the identity operator while $l_1$ and $l_2$ specify the components of the vector $\bm{R}$ along which we derive. The proofs of these equations can be found in the supplementary material. 
The non-adiabatic WAHTOR is based on the optimization of wave function, i.e. the execution of a VQE, and of the Hamiltonian. 
More specifically once it has been defined the initial Hamiltonian $H$, a VQE optimization of the ansatz parameters $\bm{\theta}$ is performed. Then the minimum of $E(\bm{R},\bm{\theta})$ with respect to $\bm{R}$ is found by exploiting one of the methods illustrated in section \ref{computational_details}, resulting in a $\hat{U}_1$ operator that is used to define the new Hamiltonian $\hat{U}_1^{\dag}H\hat{U}_1$ that is taken into account to perform a new VQE. The procedure is repeated until the energy difference between two consecutive VQEs is below the chosen threshold. At each step $i$ in which an operator $\hat{U}_i$ is generated, the Hamiltonian is updated and the converged one is $\hat{U}^{\dag}H\hat{U}$, where $\hat{U}=\hat{U}_1\hat{U}_2\hat{U}_3\dots$. 
The converged quantum state $\ket{\psi(\bm{\theta}_{opt})}$ minimizes the expectation value of $\hat{U}^{\dag}H\hat{U}$. At the same time, $\hat{U}\ket{\psi(\bm{\theta}_{opt})}$ can be seen as the state that minimizes $H$ and, as shown above, this one can be represented by the same quantum circuit of $\ket{\psi(\bm{\theta}_{opt})}$ if the qubits map the fermionic modes $\{b^{\dag}\}$ instead of $\{a^{\dag}\}$.  
In figure \ref{fig:scheme} the non-adiabatic and adiabatic strategies for the WAHTOR algorithm are reported. 
We stress the fact that, for the adiabatic version in each Hamiltonian optimization step, we define an energy functional and exploit the Hellmann-Feynman theorem to calculate analytically the first derivatives. For the non-adiabatic method, since we are considering an energy function, we can calculate the derivatives of each order in each Hamiltonian optimization step. Obviously, the first derivatives are determined in the same manner for both strategies. 
From a computational point of view, the VQE algorithm involves evaluating the Pauli strings that make up the Hamiltonian at each step. On the other hand, when computing derivatives for Hamiltonian optimization, the wave function has fixed parameters, resulting in consistent Pauli strings for the derivative operators.
As a result, the Pauli strings of the derivatives are computed only during the first time that they appear during the optimization and are not recalculated thereafter.
It is worth noting that a significant portion of the Pauli strings composing the derivatives are identical to those of the Hamiltonian. Instead of recomputing them, we utilize the Pauli strings evaluated during the last optimization step of the VQE algorithm.

\subsection{Quantum chemistry and Hubbard model results}
\label{results}
In this subsection, we report the results obtained by applying the optimization strategies illustrated above to the considered benchmarking systems, ranging from 8 to 12 qubits.
\begin{figure}[h]
\includegraphics[width=0.8\linewidth]{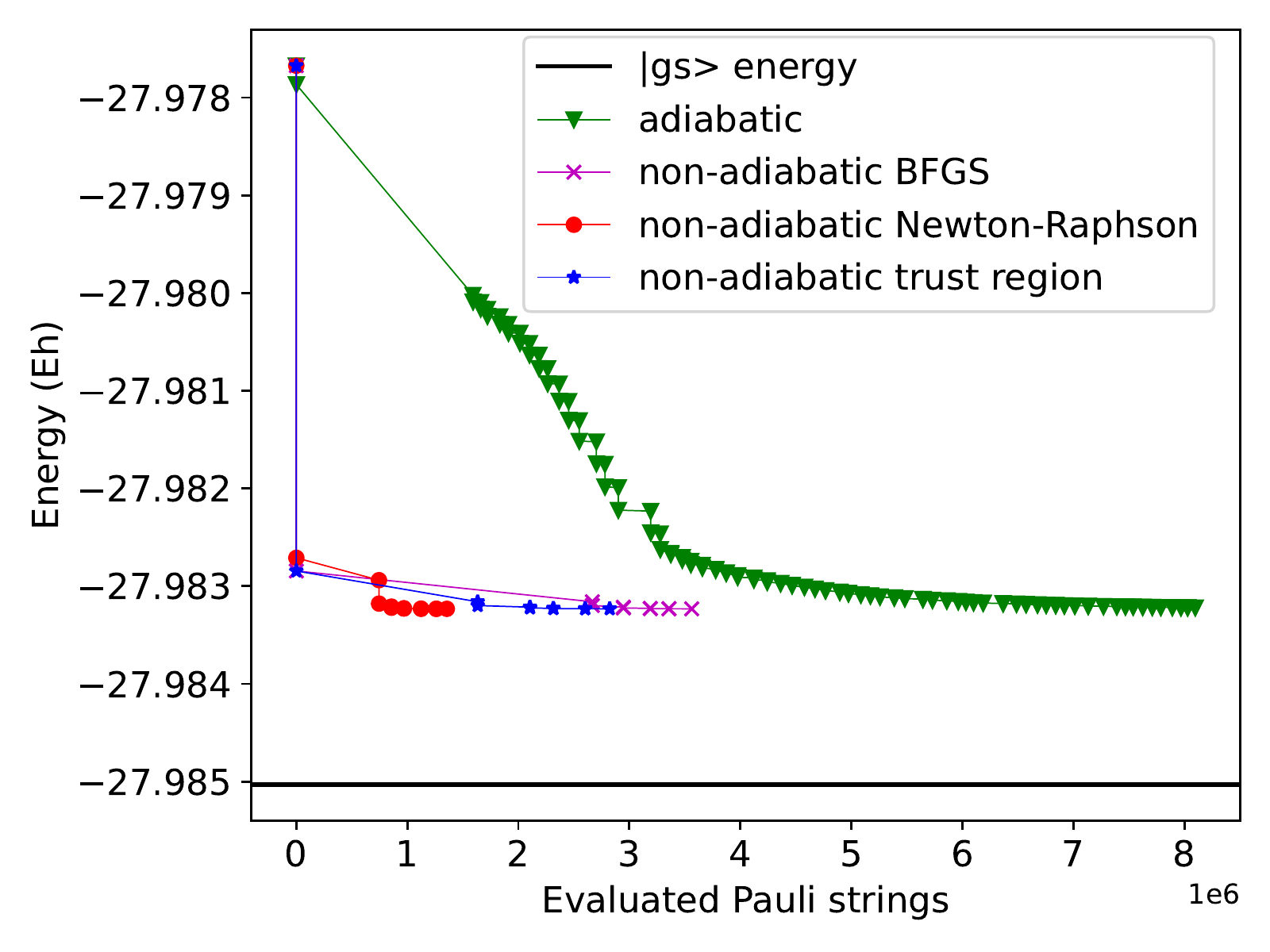}
\caption{Performance of the different optimization strategies for the $HF$ molecule. The obtained energies are reported as a function of the number of evaluated Pauli strings.}
\label{fig:HF}
\end{figure}
Figure \ref{fig:HF} shows the energy as a function of the evaluated Pauli strings for the $HF$ molecule. All the optimizers start from the same VQE state corresponding approximately to the energy of -27.978 Hartree.  Then a new Hamiltonian is generated through the calculation or the estimation of the required derivatives, giving rise to a new point on each curve.
Then a new step of optimization for the ansatz parameters is performed followed by a new generation of the Hamiltonian, as described in the scheme in figure \ref{fig:scheme}. The procedure is repeated until convergence is reached. The curves are built by reporting the energy values obtained in each step and all the optimization strategies converge approximately to the same energy value.
As can be observed in figure \ref{fig:HF}, during the derivatives calculation steps, the number of evaluated Pauli strings is negligible with respect to the ones evaluated during the ansatz parameters optimization steps. Moreover, the Pauli strings that must be evaluated to determine the derivatives are partially included in the set corresponding to the Hamiltonian. Thus their values have been calculated only in the last step of the last ansatz optimization procedure.
In addition, for the non-adiabatic methods, the expectation values of the strings can be used in the same Hamiltonian optimization, until the ansatz parameters are changed. Figure \ref{fig:HF} shows that the non-adiabatic Newton-Raphson method requires one more VQE than the other non-adiabatic methods to converge, but the corresponding Hamiltonians are composed of a smaller number of Pauli strings with respect to the BFGS and the trust region optimization strategies. 
As a consequence, the non-adiabatic Newton-Raphson optimization method results to be the most efficient one. On the contrary, the adiabatic steepest descent is the most expensive optimization procedure, as expected.
\begin{figure}[h]
\includegraphics[width=0.8\linewidth]{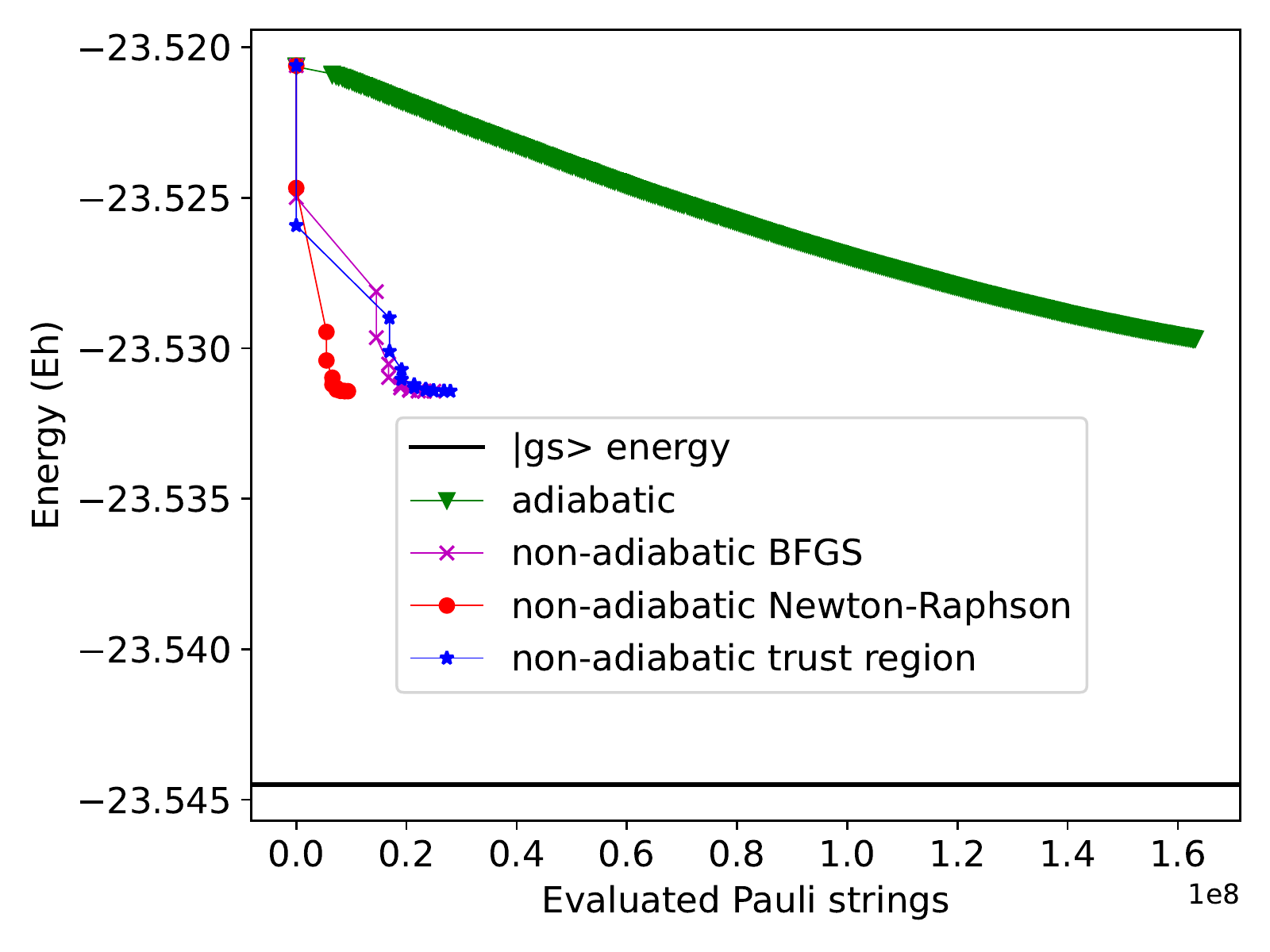}
\caption{Performance of the different optimization strategies for the $H_2O$ molecule. The obtained energies are reported as a function of the evaluated Pauli strings.}
\label{fig:H2O}
\end{figure}
\begin{figure}[h]
\includegraphics[width=0.8\linewidth]{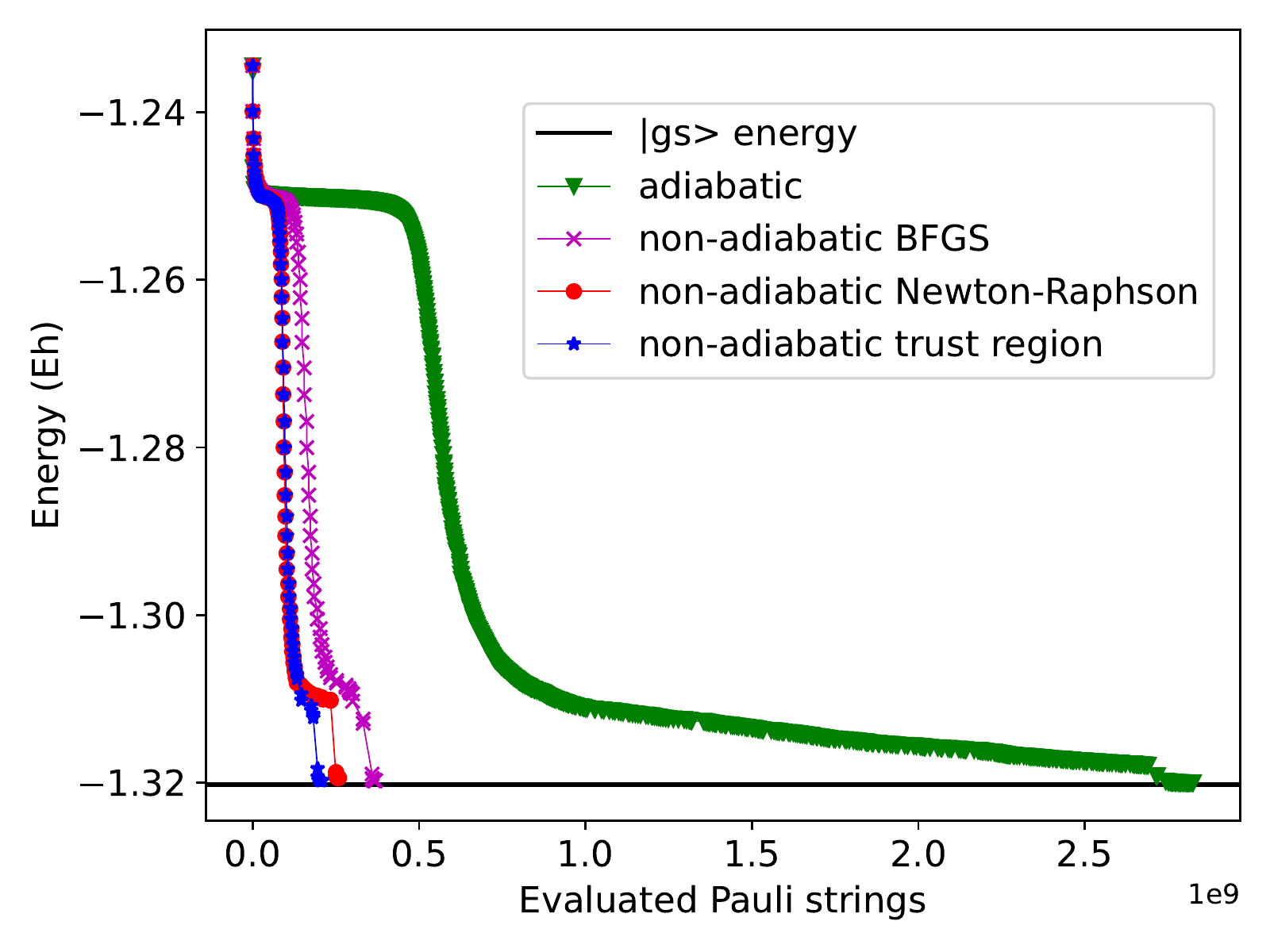}
\caption{Performance of the different optimization strategies for the Hubbard system. The obtained energies are reported as a function of the evaluated Pauli strings.}
\label{fig:Hubbard}
\end{figure}
The same considerations can be inferred for the $H_2O$ molecular system by considering figure \ref{fig:H2O}. In this case, the adiabatic steepest descent method not only requires much more QPU resources to be performed but also converges to higher energy with respect to the other optimization strategies. Even in this case, the non-adiabatic Newton-Raphson method is the most efficient strategy and it requires one less VQE than the other non-adiabatic strategies. Finally, in figure \ref{fig:Hubbard}, we report the results obtained for the 4-sites Hubbard model in the half-filling regime. The black curve in the figure refers to the exact lowest energy eigenstate. In this case, the optimization curves are substantially different from the ones obtained for the molecules due to the presence of a region with flat energy gradients. Nevertheless, the non-adiabatic schemes quickly get out of the plateau, with the non-adiabatic trust region method resulting to be the most efficient, in contrast to what is observed in molecules.

\section{CONCLUSIONS}
\label{section:conclusions}
In this work, we illustrated the non-adiabatic version of the WAHTOR algorithm, firstly developed by Ratini et al. \cite{Ratini2022}. 
We compared the efficiency of three different implementations of the non-adiabatic version of the algorithm with the adiabatic one on three benchmark systems corresponding to different cost functions and qubits numbers. 
We showed that the QPU resources spent during the energy optimization are significantly reduced in all cases with respect to the adiabatic algorithm.  
This happens because the number of VQEs steps is significantly reduced during the Hamiltonian optimization, and these steps are quite demanding in terms of QPU resources. 

Moreover, the resources are further reduced considering that some of the Pauli strings composing the derivatives of the cost function with respect to the Hamiltonian parameters have been evaluated in the last step on the last VQE. Each Pauli string is just measured one time until the ansatz parameters change, increasing the efficiency of the algorithm with this post-processing procedure. 
In particular, the non-adiabatic Newton-Raphson method was the cheapest algorithm for molecules whereas the non-adiabatic trust region was the most efficient for the Hubbard system.
Both methods require the calculation of the first and second-order derivatives, for which the analytic expression can be obtained using equations \ref{eq:dH}, \ref{eq:first_H} and \ref{eq:second_H}. The non-adiabatic BFGS method does not require the calculation of the derivatives and never results to be the most efficient, but the estimation of the second-order derivatives is enough to reduce significantly the quantum resources requested with respect to the adiabatic method. We infer that the hessian of the cost function, or its approximation, is necessary to reach our scope. Moreover, as shown in the supplementary material, we have an analytic expression of the derivative of every order for $\bm{R=0}$. They can be used to implement a non-adiabatic strategy that uses higher-order derivatives, such as the Halley method, that could give an advantage for systems that remain trapped in local minima with the strategies tested up to now. 
\cite{Gander1985,Gnang2018} 

\section{Acknowledgments}
We acknowledge fundings from Ministero dell’Istruzione dell’Università e della Ricerca (PON R \& I 2014-2020) and ISCRA C grant number HP10CQYD2V (CINECA).
\section{Competing interests}
The authors declare no competing financial interest.

\bibliographystyle{ieeetr} 
\bibliography{BIB.bib} 

\pagebreak

\begin{center}
\textbf{\large Supplementary note}
\end{center}
Here we show how to calculate the derivatives of the one and two-body integrals with respect to the Hamiltonian parameters.
We start defining the vector $\bm{R}\in\mathbb{R}^{d^2}$ and the vector $\bm{T}$ of the generators of the Hermitian matrices of dimension $d^2$. The equation of the n-th derivative of the energy with respect to the variables with indices $l_1,l_2,...,l_n$ is
\begin{equation}
\begin{split}
\frac{d^n E(\bm{R},\bm{\theta})}{d R_{l_1},..,dR_{l_n}}=\sum_{i,j}\frac{d^n h(\bm{R})_{ij}}{d R_{l_1},..,dR_{l_n}}\bra{\Psi(\bm{\theta})}a^\dag_i a_j\ket{\Psi(\bm{\theta})}+\\ +\frac{1}{2}\sum_{c,d,e,f}\frac{d^n g(\bm{R})_{cdef}}{d R_{l_1},..,dR_{l_n}}\bra{\Psi(\bm{\theta})}a^\dag_c a^\dag_d a_e a_f\ket{\Psi(\bm{\theta})}
\end{split}
\end{equation}
We note that to calculate the derivatives of each order of the energy with respect to the Hamiltonian parameters, we just need to calculate the derivatives of the one-body and two-body integrals
\begin{equation}
h(\bm{R})=e^{-i\bm{R}\cdot\bm{T}}h_{HF}e^{i\bm{R}\cdot\bm{T}}
\end{equation} 
and
\begin{equation}
g(\bm{R})=e^{-i\bm{R}\cdot\bm{T}}\otimes e^{-i\bm{R}\cdot\bm{T}}g_{HF}e^{i\bm{R}\cdot\bm{T}}\otimes e^{i\bm{R}\cdot\bm{T}}
\end{equation}
where $h_{HF}$ and $g_{HF}$ are the one and the two body integrals expressed in Hartree-Fock basis, respectively. For this purpose, we can use the following lemma of the Baker–Campbell–Hausdorff formula
\begin{equation}
e^{A}oe^{-A}=\sum_{k=0}^{\infty}\frac{[(A)^k,o]}{k!}
\end{equation}
where $[(A)^k,o] = \underbrace{[A, .., [A[A,}_\text{k times}  o]]...]$ is the iterated commutator for the two operators $A$ and $o$, that can be written as
\begin{equation}
[(A)^k,o]=\sum_{B^k=Perm\{A^k\}}[B^k_1,[B^k_2,[B^k_3,...,[B^k_k,o]...]]]
\end{equation}
where $B^k$ is a set composed of $k$ objects that, in this case, are the same operator $A$ repeated $k$ times. Now suppose that $A$ is linear in $\bm{R}$, so, defining $e^{A(\bm{R})}oe^{-A(\bm{R})}=o(\bm{R})$ it easy to show that
\begin{equation}
\frac{d^no(\bm{R})}{dR_{l_1},..,dR_{l_n}}=\sum_{k=0,n\le k}^{\infty}\sum_{B^k}\frac{1}{k!}[B^k_1,[B^k_2,[B^k_3,...,[B^k_k,o]...]]]
\end{equation}
where 
\begin{equation}
B^k=Perm\{A^{k-n},\frac{dA}{dR_{l_1}},..,\frac{dA}{dR_{l_n}}\}
\end{equation}
The derivatives are easy to evaluate in $\bm{R}=\bm{0}$ because, in this case, just the term with $k=n$ survive in the summation
\begin{equation}
\frac{d^no(\bm{R})}{dR_{l_1},..,dR_{l_n}}|_{\bm{R}=\bm{0}}=\sum_{B^n}\frac{1}{n!}[B^n_1,[B^n_2,[B^n_3,...,[B^n_n,o]...]]]
\end{equation}
with
\begin{equation}
B^n=Perm\{\frac{dA}{dR_{l_1}},..,\frac{dA}{dR_{l_n}}\}
\end{equation}

This equation can be written recursively as
\begin{equation}
\frac{d^no(\bm{R})}{dR_{l_1},..,dR_{l_n}}|_{\bm{R}=\bm{0}}=\sum_{j=Perm\{l\}}\frac{1}{n}\left[\frac{dA}{dR_{j_1}},\frac{d^{n-1}o(\bm{R})}{dR_{j_2},..,dR_{j_{n}}}|_{\bm{R}=\bm{0}}\right]
\label{derivatives}
\end{equation}

For the one body term, we have that $A=-i\bm{R}\cdot\bm{T}$ and $o=h_{HF}$, so 
\begin{equation}
\frac{d^nh(\bm{R})}{dR_{l_1},..,dR_{l_n}}|_{\bm{R}=\bm{0}}=\sum_{j=Perm\{l\}}\frac{1}{n}\left[-iT_{j_1},\frac{d^{n-1}h(\bm{R})}{dR_{j_2},..,dR_{j_{n}}}|_{\bm{R}=\bm{0}}\right]
\end{equation}
For the two body term, we observe that 
\begin{equation}
e^{i\bm{R}\cdot\bm{T}}\otimes e^{i\bm{R}\cdot\bm{T}}=(e^{i\bm{R}\cdot\bm{T}}\otimes I)(I\otimes e^{i\bm{R}\cdot\bm{T}})=e^{i\bm{R}\cdot\bm{T}\otimes I}e^{I\otimes i\bm{R}\cdot\bm{T}}=e^{(i\bm{R}\cdot\bm{T}\otimes I+I\otimes i\bm{R}\cdot\bm{T})}
\end{equation}
where $I$ is the identity operator and the last equality is true because of the Baker–Campbell–Hausdorff formula for commutating operators. So we can write the two body integrals in the following manner
\begin{equation}
g(\bm{R})=e^{(-i\bm{R}\cdot\bm{T}\otimes I-I\otimes i\bm{R}\cdot\bm{T})}g_{HF}e^{(i\bm{R}\cdot\bm{T}\otimes I+I\otimes i\bm{R}\cdot\bm{T})}
\end{equation}
Now, using the equation \ref{derivatives} and imposing $A=-i\bm{R}\cdot\bm{T}\otimes I-I\otimes i\bm{R}\cdot\bm{T}$ and $o=h_{HF}$, we obtain a formula similar to that shown for the one body term
\begin{equation}
\frac{d^ng(\bm{R})}{dR_{l_1},..,dR_{l_n}}|_{\bm{R}=\bm{0}}=\sum_{j=Perm\{l\}}\frac{1}{n}\left[-i(T_{j_1}\otimes I+I\otimes T_{j_1}),\frac{d^{n-1}g(\bm{R})}{dR_{j_2},..,dR_{j_{n}}}|_{\bm{R}=\bm{0}}\right]
\end{equation}
Finally, we obtain the first derivatives
\begin{equation}
\begin{split}
& \frac{d h(\bm{R})}{dR_l}|_{\bm{R}=\bm{0}}=-i[T_{l},h_{HF}] \\
& \frac{d g(\bm{R})}{dR_l}|_{\bm{R}=\bm{0}}=-i[T_l\otimes I+I\otimes T_l,g_{HF}]
\end{split}
\end{equation}
and the second ones
\begin{equation}
\begin{split}
& \frac{d^2 h(\bm{R})}{dR_{l_1}dR_{l_2}}|_{\bm{R}=\bm{0}}=-\frac{1}{2}([T_{l_1},[T_{l_2},h_{HF}]]+[T_{l_2},[T_{l_1},h_{HF}]]) \\
& \frac{d^2 g(\bm{R})}{dR_{l_1}dR_{l_2}}|_{\bm{R}=\bm{0}}=-\frac{1}{2}([T_{l_1}\otimes I+I\otimes T_{l_1},[T_{l_2}\otimes I+I\otimes T_{l_2},g_{HF}]]+\\
& +[T_{l_2}\otimes I+I\otimes T_{l_2},[T_{l_1}\otimes I+I\otimes T_{l_1},g_{HF}]])
\end{split}
\end{equation}

\end{document}